\documentclass[12pt,a4paper]{article}
\usepackage{graphicx}
\usepackage{wrapfig}
\begin{document}
\textwidth=135mm
 \textheight=200mm
\begin{center}
{\bfseries Measurements of Forward Jet Production in Polarized pp Collisions at
 $\sqrt{s}=500$~GeV
}
\vskip 5mm
L.~Nogach, for the A$_N$DY collaboration
\vskip 5mm
{\small {\it Institute of High Energy Physics, Protvino, Russia
}}
\end{center}
\vskip 5mm
\centerline{\bf Abstract}
{\small
The A$_N$DY project at RHIC was proposed to measure the analyzing power for
Drell-Yan  production. Test runs took place during polarized proton operations 
of RHIC in 2011 and 2012 with a model of the A$_N$DY apparatus in place. 
In total, an integrated luminosity of 9 $pb^{-1}$ with beam polarization of 
50\% was sampled. The primary detector components were a hadron calorimeter 
(HCal) that spanned the pseudorapidity interval $2.4 < \eta < 4.0$ and a small 
electromagnetic calorimeter (ECal). Basic goals for A$_N$DY test running were 
to establish the impact of a third interaction region on RHIC performance and 
to demonstrate HCal calibration. Energy scale of HCal was established using 
neutral pion reconstruction and checked with hadronic response. In addition, 
data with a trigger based on HCal energy sum were taken to study jet events. 
First measurements of analyzing power in the forward jet production 
are reported.
}
\vskip 10mm

\section{\label{sec:ln_intro}Introduction}
~~~
\vspace{-7mm}

Large transverse single spin asymmetries (SSA) measured in inclusive pion 
production in $pp$-collisions \cite{ln_e704,ln_fpd} have stimulated significant 
theory development to understand the spin structure of the proton. However, 
this process is not easy to describe due to contributions from several 
mechanisms related to initial and final state interactions. Simpler processes 
from the theoretical point of view, such as Drell-Yan, prompt photon or jet 
production, should be considered to disentangle the different mechanisms. 
In particular, inclusive jet production is of interest since it has no 
final state (Collins effect) contribution, and arises only from the Sivers
effect. From naive expectations, jet SSA should be small because jets 
integrate over charged and neutral pions, and opposite sign asymmetry 
for $\pi^+$ and $\pi^-$ leads to cancellations. Theoretical models also 
expect jet SSA in the forward region~to~be~small~\cite{ln_dAlesio,ln_Kang}. 

First measurements of forward jet production in polarized $pp$-collisions 
were performed at RHIC 2 o'clock interaction region (IP2) during two 
 A$_N$DY test runs: at $\sqrt{s}=500$~GeV in 2011 and $\sqrt{s}=510$~GeV 
in 2012. The A$_N$DY setup in 2011 run is described in 
\cite{ln_andy-dspin2011}, and included zero-degree calorimetry (ZDC) for 
luminosity monitoring and a check of polarization direction. For 2012 run 
HCal Left/Right modules were modified into an annular detector of 
20$\times$12 cells with a 2$\times$2 hole for the beam pipe, and ECal was 
put on rails so that it could be moved out not to shadow HCal for jet 
measurements. The luminosity was 2$\times$ higher compared to the 2011 run. 
Data were taken with two basic triggers: 
1) energy sum in HCal Left/Right half, excluding two outer perimeters 
to ensure that jets are contained in the detector, with the threshold 
$\sim$\,35 GeV; 2) energy sum in ECal to measure trigger bias~on~jets. 

\section{\label{sec:ln_reco}Jet reconstruction}
~~~
\vspace{-7mm}

The energy scale for HCal was set based on neutral pion calibration 
\cite{ln_andy-dis2011} with a crude adjustment for hadronic compensation
from PYTHIA/GEANT simulations: $E' = 1.12{\times}E - 0.1$~GeV, where $E$ 
is the incident energy using $\pi^0$ calibration and $E'$ for HCal cells 
is used in jet reconstruction. $E'=0.25$~GeV threshold was applied for 
a cell to be included into a jet finding. Both ECal and HCal cells were
used in the jet finding for the 2011 run data. 

Two algorithms were used for jet reconstruction. The cone jet finder starts 
from a seed (high tower in the triggering region of the detector), sums energy 
in the cone of radius $R=0.7$ in $(\eta{-}\phi)$ space (where $\eta$ is the 
pseudorapidity and $\phi$ is the azimuthal angle relative to the beam 
direction) around the high tower and defines the jet axis from energy-weighted 
${<}\eta{>}$ and ${<}\phi{>}$. Then an iterative procedure is applied until 
convergence of the jet axis: 1) sum energy in the cone of radius $R$ about 
${<}\eta_N{>}$, ${<}\phi_N{>}$ ($N$ is the iteration number); and 
2) compute energy-weighted ${<}\eta_{N+1}{>}$, ${<}\phi_{N+1}{>}$. 

The recently developed anti-k$_T$ algorithm \cite{ln_jetreco} introduces 
a distance measure and uses sequential recombination of cells (clusters) 
to form jets. For cluster pair $i$ and $j$, the distance is computed as 
$d_{ij}=\textrm{min}(k^{-2}_{T,i},k^{-2}_{T,j}){\times}(R^2_{ij}/R^2)$, where 
$k_{T,i} = E_i/\textrm{cosh}(\eta_i)$ is the transverse momentum assuming 
zero mass for the incident particle, 
$R^2_{ij}=(\eta_i - \eta_j)^2 + (\phi_i-\phi_j)^2$, and $R=0.7$ is used. 
If $d_{ij}<1/k^2_{T,j}$ for any $i$, the clusters are merged and the procedure 
is repeated, otherwise cluster $j$ is considered as a jet. 
Finally, energy and acceptance cuts were imposed to select ``good'' jets: 
$E_{jet} > 30$~GeV, $|\eta_{jet}-3.25|<0.25$, $|\phi_{jet}-\phi_{off}|<0.5$, 
where $\phi_{off}=0~(\pi)$ for the Left (Right) module. 

\begin{figure}[t]
  \includegraphics[width=0.315\textwidth]{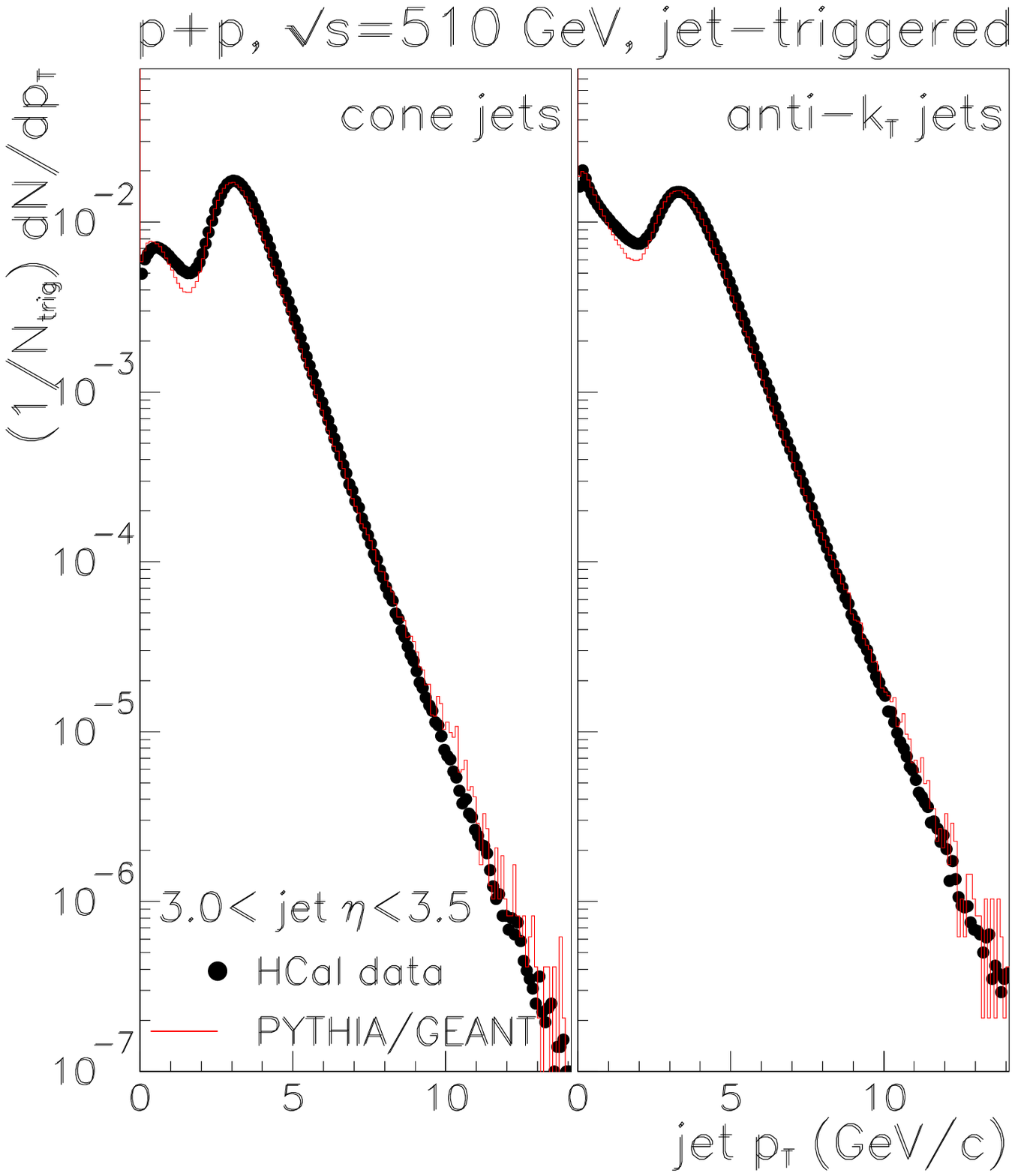}
  \includegraphics[width=0.315\textwidth]{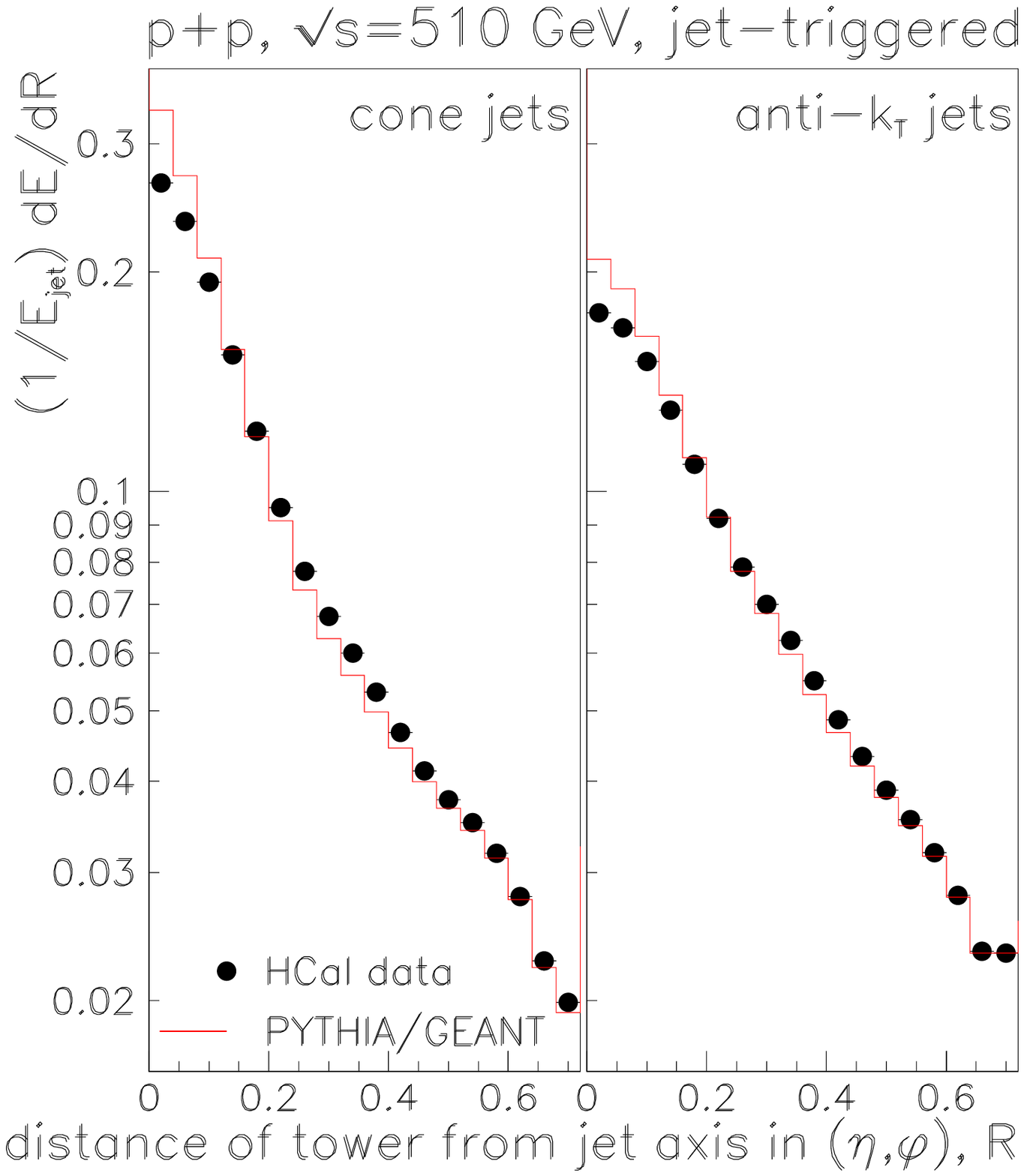}
  \includegraphics[width=0.37\textwidth]{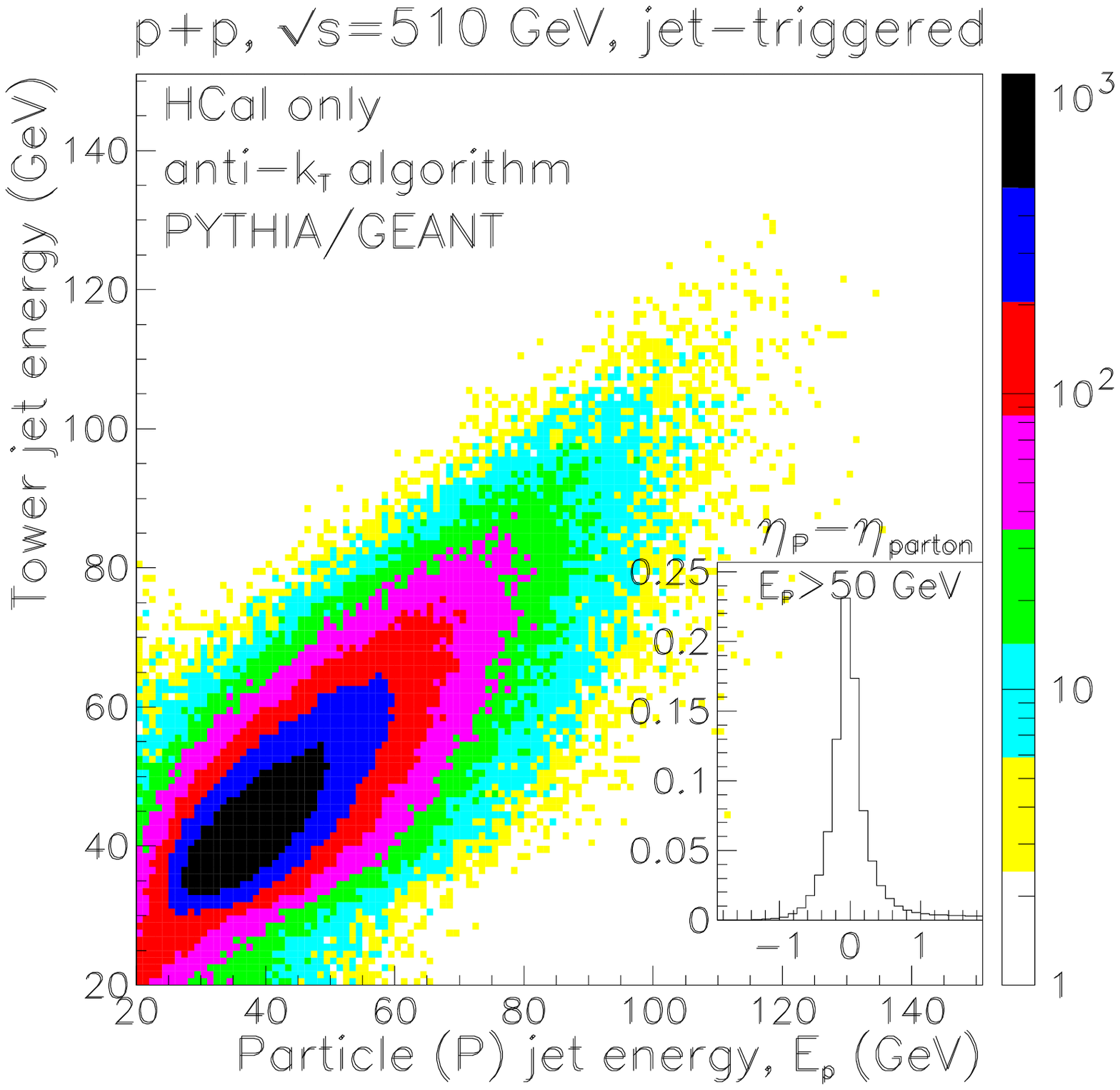} \\
  \hspace*{2.2cm}\small{(a)}\hspace{3.8cm}\small{(b)}\hspace{4cm}\small{(c)}
  \caption{\small (a) Distribution on jet transverse momentum $p_T$, 
   normalized by number of triggers. (b) Distribution of fraction 
   of energy in the jet as a function of distance in ($\eta{-}\phi$) space 
   from the jet axis (jet shape). (c) Correlation between tower jet and 
   particle jet energy. The inset shows the $\eta$-component of the direction 
   match ($\Delta\eta$) between particle jets and hard-scattered parton,
   whose direction is defined by $\eta_{parton}$, $\phi_{parton}$. 
   There is 82\% match requiring $|\Delta\eta|, |\Delta\phi|<0.8$.}
  \label{ln_fig1}
\end{figure}

Jet shape and transverse momentum distributions from the two algorithms 
are shown in Fig.\ref{ln_fig1} for the jet-triggered data and simulations. 
There is good agreement between the data and simulations above the trigger 
threshold for both algorithms. There are some quantitative differences between 
the algorithms: cone jets have more steeply falling $p_T$ distribution and 
more narrow jet shape than observed for anti-$k_T$ jets. This is likely 
related to additional ``out-of-cone'' cells acquired by the anti-$k_T$ 
algorithm. Hard-scattered partons are strongly correlated with jets, and 
the jet energy scale is checked by correlating "tower" jets reconstructed 
from PYTHIA/GEANT (i.e., full detector response) simulations versus jets 
reconstructed from particles generated by PYTHIA, as shown in 
Fig.\ref{ln_fig1}c. 

\section{\label{sec:ln_res}Jet analyzing power}
~~~
\vspace{-7mm}

The jet asymmetry was calculated from yields in the Left/Right module sorted 
by the polarization direction of the beam heading towards the detector
for $x_F >0$ and the opposite beam for $x_F<0$: \\
$\epsilon = P_{Beam}A_N = 
(\sqrt{N^{\uparrow}_L N^{\downarrow}_R} - \sqrt{N^{\uparrow}_R N^{\downarrow}_L})/
(\sqrt{N^{\uparrow}_L N^{\downarrow}_R} + \sqrt{N^{\uparrow}_R N^{\downarrow}_L}),$ 
where $N^{\uparrow(\downarrow)}_{L(R)}$ is the number of jet events in the Left 
(Right) module for the spin direction up (down), as determined from our 
measured ZDC spin asymmetries. This method relies on mirror symmetry in 
the setup geometry, and cancels systematics, such as detector and 
luminosity asymmetries, through second order. To check time-dependent
systematics, the asymmetry $\epsilon$ was computed in $x_F$ bins for each 
RHIC fill and fitted by a constant function. $\chi^2$ per degree of freedom 
from these fits were close to 1, meaning that systematic errors are small. 
Bunch shuffling, i.e. random reversing of the spin direction for half of 
the filled bunch crossings that creates effectively unpolarized collisions, 
was used as another estimate of systematics. Mean value of the asymmetry 
calculated for $\sim$100 random patterns was ${\sim}(10^{-4}{-}10^{-5})$, 
resulting in the systematic uncertainty in $A_N$ less than $2{\times}10^{-4}$.

\begin{wrapfigure}[18]{R}{6cm}
  \vspace{-7mm}
  \includegraphics[width=6cm]{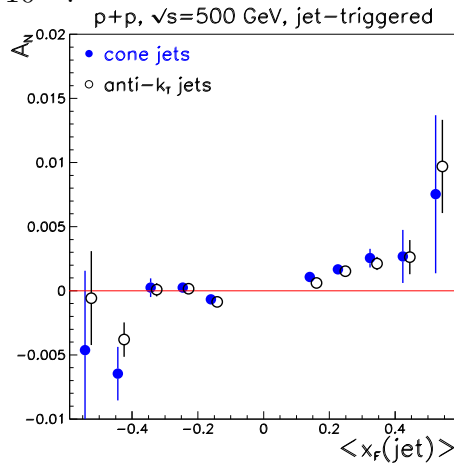}
  ~~~
  \vspace*{-8mm}
  \caption{\small Jet $A_N$ as a function of $x_F$ for jet-triggered
   events. Cone jet points are shifted from the mean $x_F$ value in the bin by 
   -0.01, and anti-$k_T$ jet points are shifted by 0.01. Error bars indicate 
   statistical errors.}
  \label{ln_fig2}
\end{wrapfigure}

The analyzing power $A_N$ was calculated using fill-averaged 
beam polarization 0.52 for both beams, and is shown in Fig.\ref{ln_fig2} 
for the cone and anti-$k_T$ jet finders. The results from the two algorithms 
are consistent within statistical errors. $A_N$ at $x_F>0$ is small 
(${\sim}10^{-3}$) and positive (up to $6\sigma$). Jet $A_N$ was also measured 
for ECal-triggered events, and turned out to be at the level (1-3)\% for 
$x_F>0$. This likely means that trigger bias imposed by electromagnetic 
calorimeter prefers jets that fragment to a hard neutral pion. 

We have found the jet analyzing power to be small and positive. Although
our measurements can help to further constrain the Sivers functions, it is
most important to measure the analyzing power for Drell-Yan production as a
test of present understanding.


\begin{thebibliography}{9}
\bibitem{ln_e704}
            \textit{Adams D.L. et al.} //
            {Phys. Lett. B. 1991. V.261. P.201;} 
            {Phys. Lett. B. 1991. V.264. P.462.}
\bibitem{ln_fpd}
            \textit{Abelev B. et al.} //
            {Phys. Rev. Lett. 2008. V.101. P.222001.} 
\bibitem{ln_dAlesio}
            \textit{D'Alesio U., Murgia F. and Pisano C.} //
            {arXiv:1011.2692.}
\bibitem{ln_Kang}
            \textit{Kang Z.-B. et al.} //
            {arXiv:1103.1591.}
\bibitem{ln_andy-dspin2011}
            \textit{Nogach L. et al.} //
            {arXiv:1112.1812.}
\bibitem{ln_andy-dis2011}
            \textit{Perkins C.} //
            {arXiv:1109.0650.}
\bibitem{ln_jetreco}
            \textit{Cacciari M., Salam G.P. and Soyez G.} //
            {JHEP 2008. V.0804. P.063.}
\end{thebibliography}
\end{document}